\documentstyle[aps,12pt]{revtex}
\begin{document}
\begin{center}
\noindent
 {\Large \bf A decommissioned LHC model magnet as an axion telescope}

\end{center}\vskip0.6cm
 
\noindent  K.~Zioutas~$^{1}$,~C.E.~Aalseth~$^{2}$,~ D.~ Abriola~$^{3}$,
 ~F.T. ~Avignone~III~$^{2}$,~R.L.~Brodzinski~$^{4}$, ~J.I.~ Collar~$^{5~*}$,
  ~R.~Creswick~$^{2}$, ~D.E. ~Di~ Gregorio~$^{3}$,~H. ~Farach~$^{2}$,
   ~ A.O. ~Gattone~$^{3}$, ~C.K. ~Gu\'erard~$^{3}$,~F. ~Hasenbalg~$^{3}$,
    ~M.~Hasinoff~$^{6}$, ~H.~Huck~$^{3}$, ~A.~ Liolios~$^{1}$, ~H.S. ~Miley~$^{4}$,
     A.~Morales~$^{7}$,~ J.~Morales~$^{7}$,~ D.~Nikas~$^{1}$, ~S. ~Nussinov~$^{8}$,
     ~A.~ Ortiz~$^{7}$, ~E.~Savvidis~$^{1}$,~ S. ~Scopel~$^{7}$, 
     ~ P. ~Sievers~$^{5}$,~J.A. ~Villar~$^{7}$,~ L. ~Walckiers~$^{5}$. 
\vskip0.5cm


\noindent
{\small \bf 1)}~~{\it Physics Department, University of Thessaloniki, 
  GR-54006  Thessaloniki,  Greece.}
\noindent
 {\small \bf 2)}~~{\it Department of Physics and Astronomy, University of South
 Carolina, Columbia, 
 
 ~SC 29208, USA.} 

\noindent
 {\small \bf 3)}~~{\it Department of Physics, TANDAR Laboratory, C.N.E.A., Buenos
 Aires, Argentina.}

\noindent
 {\small \bf 4)}~~{\it Pacific Northwest National 
 Laboratory, Richland, WA 99352, USA.}

\noindent
 {\small \bf 5)}~~{\it CERN, CH-1221 Geneva 23, Switzerland.}

\noindent{\small \bf 6)}~~{\it Department of Physics and Astronomy,
 University of British Columbia, Vancouver   
 
 ~BC V6T IZI, Canada.}

\noindent
{\small \bf 7)}~~{\it Laboratorio de Fisica Nuclear y Atlas Energias,
 Faculdad de Ciencias, Universidad 
 
 ~de Zaragoza, E50009 Zaragoza, Spain.}

\noindent
{\small \bf 8)}~~{\it Department of Physics, Tel Aviv University, Tel Aviv,
 Israel.}

\begin{center}
\vskip0.3cm
{\it       }
\vskip0.3cm
\end{center}
\noindent
{\bf Abstract.}
The 8.4 Tesla, 10 m long transverse magnetic field of a twin aperture
LHC bending magnet 
can be utilized as a macroscopic coherent solar
axion-to-photon converter.
Numerical calculations show that the integrated time of alignment with the Sun would 
be 33 days per year with the magnet on a tracking table capable of $\pm 5^o$ in 
the vertical direction and $\pm 40^o$ in the horizontal direction. The existing 
lower bound on the axion-to-photon
coupling constant can be improved  by a factor between 30 and 100 in 3 years,
 i.e., ${\it g_{a\gamma \gamma}} \lesssim 9\cdot 10^{-11}$ GeV$^{-1}$ for
 axion masses  $\lesssim$ 1 eV. This value falls within the existing 
open axion mass window. The same set-up can simultaneously search for low- and 
high-energy celestial axions, or axion-like particles, scanning 
the sky as the Earth rotates and orbits the Sun.
\vskip0.5cm
PASC numbers: 41.85.Lc, 85.25.Ly, 14.80.-j, 14.80.Mz, 92.60.Vb, 95.
\vskip0.3cm\noindent$-----------------------------$\noindent

\vskip0.1cm
$^*)$ ~ E-mail : JUAN.COLLAR@CERN.CH
\vskip0.3cm
\noindent
~Proofs should be sent to: J.I. Collar (address above), phone: +331 44278237, 
\vskip0.1cm
\noindent
~fax: +331 43542878, collar@mail.cern.ch

\newpage

\section*{1. Introduction}

QCD, the universally accepted theory of strong interactions has  one 
serious blemish, the
U(1) problem (why the $\eta '/\pi$ mass ratio is not closer to unity
\cite{wein}).
Its solution produces in turn the strong CP problem (why the neutron's electric
dipole moment is at least a factor $10^{-9}$ smaller than expected). An attractive solution
to this invokes a new U$_{PQ}$(1) symmetry, the Peccei-Quinn symmetry.
The spontaneous breaking of this new symmetry predicts the existence of a light
neutral pseudoscalar particle, the axion, which is closely related
to the neutral pion
\cite{pq,ww}.
The axion also arises in supersymmetry and superstring theories.
In addition, axions are one of
the most interesting non-baryonic candidates for the ubiquitous
dark matter (DM) 
\cite{je,ggr,ps,slcheng}.
Axions may also exist as primordial cosmic relics copiously 
produced in the early Universe. For these reasons, axions have
received much attention. More details about axions and their role in 
cosmology and astrophysics can be found in \cite{smlee,jorge}. A remarkably 
pedagogical introduction to them is given by Sikivie in \cite{jicsik}.

The search for relic axions is based mainly on the Primakoff effect
and is performed with superconducting resonant cavities
\cite{vanb,ch,io} 
or, also recently, with CERN's SMC polarized target 
\cite{yks}.
{\sf Fig.~1}, taken from ref.
\cite{yks}, shows the present status for axion searches.
The main limitation of the  sensitive experiments  described in \cite{vanb,ch,io} is
the possibility that the
 axion rest mass lies outside the resonance region of the micro-wave cavities utilized.
However, energetic axions might also be
created continuously in reactions taking place in 
 red giants, supernovae, and in particular inside our Sun, this being 
the nearest and brightest potential axion source in the
1$-$15 keV total energy range. Experiments searching for solar
axions have the main advantage of being sensitive in principle to
a very broad axion rest mass range ($m_a c^2\lesssim 10$ keV), although coherence-loss
constraints decrease  the  sensitivity for axion masses above $\sim$ 1 eV \cite{ps}.

Astrophysical considerations (cooling rates of stars), cosmological
arguments (overclosure of the Universe), and laboratory experiments
require an axion mass in the range
\cite{ggr,ps,slcheng,graw,svk,htj,mtr}
\begin{equation} 10^{-5} eV < m_ac^2 < 10^{-3} eV ,
\end{equation}
while the mass range
\begin{equation}
2~eV < m_ac^2 < 10~eV 
\end{equation}
is still possible for hadronic axions with couplings
to leptons equal to zero
\cite{ps,kep}.

The axion decay into two photons
(${\sf \it a \rightarrow \gamma \gamma}$) is the reaction mainly used to
search for axions. Inside a magnetic field, the axion
couples to a virtual photon ($\gamma_{virtual}$),
producing a real photon ($\gamma$)  via the Primakoff effect. 
This real photon can then be  detected :
\begin{equation}
a ~ +~ \gamma_{virtual} \rightleftharpoons \gamma 
\end{equation}
The energy of this photon ($\gamma$) is then equal to the axion's
total energy. The average energy of the emitted solar axions,
and therefore of the converted photons is $\sim$ 4.2 keV
\cite{kvbi}. 
In principle, reaction (3) allows  for (coherent) photon mixing with spin  
0, 1, or 2 particles (e.g. spin-0 for axion and spin-2 for the
{\sf \it graviton}
\cite{grls}). An improvement in the axion-to-photon coupling limit
implies an equivalent improvement in the axion mass limit; these
quantities are related by
\cite{emrt}
\begin{equation}
g_{a\gamma \gamma}
\equiv M^{-1} \approx 10^{-10} \left[ \frac{m_a}{1 eV}\right] GeV^{-1} ,
\end{equation}
where $M$ is the corresponding mass scale of the spontaneously broken new
symmetry (U$_{PQ}$(1)). Models of hadronic axions with enhanced or supressed
photon couplings are easily constructed \cite{kapl}; here  
 the factor in front 
of $m_a$ in Eq. (4) is chosen to vary within an indicative range $10^{-9}-10^{-11}$, to form 
the broad band labeled "axion models" in {\sf Figs.~ 1,3}.

In this paper it is shown that the first twin-aperture superconducting
LHC prototype magnets at CERN have an interesting application as a solar axion detector,
without interfering with the LHC
construction (these are test magnets and will become available for this 
independent experiment at the end of 1998. A suitable experimental hall 
to house the set-up will be required).
The general experimental concept is to have the magnet aligned with the Sun
for a period as long as possible, using an active tracking system; the
Primakoff converted photons, emitted in the direction of the incoming
axion and carrying its original energy and momentum, are then detected
by ultralow-background detectors at the end of the magnetic pipe lines.
A hypothetical axion signal is then expected only during the times of
solar alignment with the magnet axis, providing a unique signature.
 The suggested axion telescope can also discover or
exclude  other hypothetical (axion-like) celestial particles
with similar couplings, with  energies above $\sim$ few keV.

\section*{2. Axion-to-photon conversion}

An experimental limit for the axion-to-photon coupling inside a
magnetic field  was obtained in 1992 by the pioneering axion
experiment of the Rochester-BNL-FNAL
collaboration
\cite{dml} :
\begin{equation}
g_{a\gamma \gamma}<3.6\cdot 10^{-9}GeV^{-1}~for~m_a<0.03~eV~ 
and~ g_{a\gamma \gamma} \lesssim 7.7\cdot 10^{-9}GeV^{-1}~for~
0.03<m_a \lesssim 0.1~eV\end{equation}
Most of the theoretical relations and the expected experimental 
parameters of relevance are taken from refs.
\cite{kvbi,grls,dml,fh}.
The magnetic field strength ({\sf B}) and its length ({\sf L}) are the 
fundamental parameters in the calculation of the performance of a detector to
be used as a coherent axion-to-photon converter. The probability of
detecting a photon in the $\sim~1-15$ keV region, per solar axion
entering one of the magnetic field pipes, is 
\begin{equation}
P_{a\rightarrow \gamma} = \left[\frac{ {\sf B \cdot L} }{2M}\right]^2 ,
\end{equation} 
assuming 100$\%$ detection efficiency for the conversion x-rays.

\noindent
The conversion probability is :
\begin{equation}
P_{a\rightarrow \gamma} \approx 
1.8\times 10^{-17}\times\left(\left[\frac{{\sf B}}{8.4T}\right]^2
\cdot\left[\frac{{\sf L}}{10m}\right]^2\cdot\left[\frac{10^{10}~
GeV}{M}\right]^2\right)
\end{equation}
(with $g_{a\gamma \gamma}\equiv 1/M$).

\noindent
On Earth, the total
solar axion flux $\Phi_a$ is
\cite{kvbi} :
\begin{equation}
\Phi_a \approx 3.5\cdot 10^{11} 
\left[\frac{10^{10}~GeV}{M}\right]^2 /~cm^2\cdot s
\end{equation}  
It follows that
 the integrated axion flux~ $\Phi_a^{LHC}$~  crossing an LHC
bending magnet (with a cross-section area $F\approx 2\times 14~cm^2$), while being
aligned with the Sun's core, is :
\begin{equation}
\Phi_a^{LHC} \approx ~8.6\times 10^{17} \left[\frac{10^{10}~GeV}{M}\right]^2 axions/d .
\end{equation}   
Combining the axion conversion efficiency (Eq. (7), evaluated at the indicated 
fiducial values
of {\sf B},{\sf L}) with
the axion fluence (Eq. (9)), the estimated number
(R) of x-rays due to the converted axions for a
 solar alignment 
time of 24 $h$ within 1 year, attainable with just a $\pm~1.5^o$ tracking along the
horizontal and vertical axis ({\sf Fig.~ 2}), is
\begin{equation}
R\approx 1.8\cdot 10^{-17}\times 8.6\cdot 10^{17}
\cdot\left[\frac{10^{10}GeV}{M}\right]^4=15~events/(24 h~per~year),
\end{equation}
given $M~=~10^{10}~GeV$. Taking into account 
the estimated  background in the x-ray detectors
(Eq. (14)-(15)), discussed below, a value for the coupling constant 
$g_{a\gamma \gamma}>10^{-10}~GeV^{-1}$ is then
expected to be rejected at the $3\sigma$ level 
within the first 2 years, even with this modest
$\pm~1.5^o$ tracking. This limit would be better  by a factor
 $\sim$ 30  than that obtained in ref. \cite{dml}.
In order to show the maximum improvement reachable with only one LHC
model bending magnet, these figures can be scaled up.
Let us take 2 years of measuring time, $\pm 40^o$ horizontal and 
$\pm 5^o$ vertical tracking (=33 days of integrated solar-alignment 
time per year, {\sf Fig.~ 2}).
Relation (10) then yields 
\begin{equation}
R\approx 195~events/(66~days~in~2~years) ,
\end{equation}
when $M~=~1.51\cdot10^{10}~GeV$  ($g_{a\gamma\gamma}~=~6.6\cdot 10^{-11}~GeV^{-1}$) ({\sf Fig.~ 3}).
This rate is roughly three times the statistical fluctuation in the background estimated in Eq. (15). 
Therefore, the new limit on the axion-to-photon coupling constant, at the
$\sim 3\sigma$ level, and for comparison with ref.\cite{dml}, would be 
\begin{equation}
g_{a\gamma \gamma} < 6.6\cdot 10^{-11}~GeV^{-1}.
\end{equation}
This represents an improvement by a factor $\sim$ 55. Similarly, considering one single 
14 m bending magnet or two 10 m  magnets in series, the expected limits for the
 coupling constant $g_{a\gamma \gamma}$ are $5.2\cdot 10^{-11}~GeV^{-1}$ and
  $4.6\cdot 10^{-11}~GeV^{-1}$, respectively. The possible improvement is then
   a factor of 70-80. The $M^{-4}$ dependence
of the (solar) axion-to-photon conversion rate
(Equation (10)) makes an experiment with
further improvement rather difficult.

\noindent
{\bf \underline{Coherence}} :
For massive axions, in order to fulfill the coherence relation
(6), i.e. to avoid deconstructive axion-photon interference,
the magnetic field length ($L$) must be
\cite{dml}
 :\begin{equation}
L~ <~ \frac{(2\pi \hbar c)\cdot (\hbar \omega)}
{m_a^2c^4}
\end{equation}
For example, a coherence length of $L=10-20$ m in vacuum requires
$m_a\leq 0.01$ eV for a photon energy $\hbar \omega$$\approx$5 keV;
this condition is fulfilled when the x-ray detector is placed outside the magnet
(see ref. \cite{dml}).
Note that according to relation (13),  axions of higher energy 
have an accordingly longer coherent conversion pathlength. 
Thus, 511 keV and $\sim$ 60 MeV axions, which could be produced during
positron annihilation and supernovae explosions 
\cite{emrt},
can be similarly detected with $L$ $\sim$ 10 m magnets even if they have a rest mass of up to 
0.1 or 1 eV, respectively.  In other words, such a powerful
axion-to-photon converter (combined with
 a high-efficiency, low-background  detector),
can be used to simultaneously search for low- as well as high-energy
 celestial axions;
the active anti-Compton shielding (see below) can be used for this purpose too.
Due to the Earth's motion almost the whole sky can be scanned 
progressively ({\sf Fig.~4}). 

Coherence can be restored for a solar axion rest mass of up to 
$\sim$ 1 eV by filling the magnetic conversion region with buffer gas \cite{kvbi},
e.g. He gas at a pressure up to 10 atm, this being the limiting value since the absorption
length for  $\sim$ 4 keV photons then approaches 10 m, the magnet length.
For an appropriate gas pressure, the `dressed' photons inside the buffer gas, coming
from the converted axions, acquire an effective mass ($m_{\gamma}$) whose
wavelength can match the axion's wavelength, thus preventing 
deconstructive interference, i.e., coherence is preserved for a narrow mass
window. By limiting ourselves to $<$ 10 atm, no complication is expected due to photon interaction with
the buffer gas, whose pressure should be varied in appropriate steps
(after the first $\sim$ 2 years of measurement in vacuum), in order
to cover axion masses above $\sim$ 0.01 eV.
In this manner the expected sensitivity on the axion-to-photon coupling constant can reach the
theoretically favoured regions above $\sim$ 0.01 eV ({\sf Fig. 1,3}).
The operating pressure (at $300^o$ K) is $P[atm]\approx 15\cdot m_{\gamma}[eV]$
for a chosen `photon mass', $m_{\gamma}$ \cite{kvbi}. This type of measurements
will require an additional 1-2 years to be completed, in order to measure several
points in the range $0.02 \leq m_a \leq 1$ eV and  $g_{a\gamma \gamma}< 10^{-10}
GeV^{-1}$ (the Red Giant limit).
However, even with vacuum in the magnetic field pipes,
the axion rest mass range that
can be investigated still fully covers the allowed region given in (1).

\section*{3. The photon detector}

In order to minimize Compton scattering from higher energy background photons,
 a commercial detector consisting of a thin Ge crystal with an integral
 Compton suppression crystal should be used. A
 low-background electro-formed Cu cryostat and 
  $\sim$ 20 cm  Pb shielding 
 surrounded with a 10 cm plastic scintillator, to veto cosmic-rays, will
 further suppress the background. Additional moderator outside the plastic
 would be desirable as well.
 Special precautions against low-energy microphonics must be taken using well-known
 discrimination techniques based on pulse-shape analysis
 \cite{morales}. Thin n-type
 Ge detectors with a Be window (53 mm in diameter, 50 g in weight) manufactured
 to our specifications are being considered. The expected energy threshold
 is $\sim$ 600 eV and the low-energy resolution is  FWHM $\approx$ 400 eV.
 Similar low-background experiments have been performed at ground level by members of
 this collaboration \cite{miley}.

\noindent
{\bf \underline{Background estimation.}}~ The achievable background ($N$) at sea level 
for the two Ge x-ray detectors (50 grams each), one at each end of the magnet,
is
\cite{miley,heusser} :
\begin{equation}
N\approx 50-80~counts/d\cdot kg\cdot keV~\approx~65~counts/
d\cdot 2\times 50g \cdot 10keV
\end{equation}
(10 keV is the width of the solar axion spectrum).

The experimentally determined background level, for the
$\sim$ 66 days of integrated exposure to the Sun during the first 2 years, will be
\begin{equation}
N~\approx~ 
(4290~\pm~65)~counts~/~66d\cdot 2\times 50g \cdot 10keV.
\end{equation}
However, it is the background fluctuation ($\pm 65$) that imposes a limit
on the maximum allowable number 
of axion events (Equations  (10) and (11)),
and not the background level itself. The uncertainty in the average background 
can be made negligible by running the detectors (with the magnet switched-off) during the 
times of solar misalignment. There is no immediate advantage in bringing the set-up
 to an underground location, since the further reduction in background (a factor 
 of $\sim$20 \cite{we}) translates into an increase in $g_{a\gamma \gamma}$ sensitivity 
 of only $\sim$1.5 (Eq. 10, 11). This relatively weak dependence ($\sim$  $N^{1/8}$) of the achievable limits
 on the detector background may make of PIN diodes an inexpensive 
 alternative to be considered \cite{mino1,mino2}.
 
\section*{4. The Magnet}

{\bf 4.1 ~The existing set-up}

\noindent
First generation twin aperture superconducting LHC magnets have been built and tested
\cite{m1}. Their nominal field is 8.4 T, but they routinely reach fields
between 9 and 9.5 T. Their effective magnetic length is 9.25 m. These prototype magnets
are not bent to cope with the LHC radius of curvature. They have straight
cold bores of 42.5 mm aperture. All these magnets have been tested and measured 
magnetically on a bench in the SM18
facility at CERN
\cite{m2}. Three of the seven magnets tested underwent
a thorough life test.
A 30 m long girder is used to support the magnet under test together with the MFB
(Magnet Feed Box) housing all the cryogenic and electrical connections needed
for the tests in superfluid helium.  The girder can be presently
tilted about two rotation axes to simulate the possible inclinations of the magnets
in the LEP/LHC tunnel ($\pm ~1.5 \%$ of slope). 
The magnetic measurements were performed with rotating coils at room temperature
by sliding warm fingers inside. These anticryostats have a 35 mm  aperture and a
length of 13.8 m to go through the MFB and have been used opened at both ends
or closed  and pumped down.

\vskip0.5cm

\noindent
{\bf 4.2 ~ Necessary improvements}

\noindent
The stroke of the jack can be increased, in order to provide an inclination of the
girder to $\pm 10^o$. The shear stress so produced by the magnet's weight in
the cryostat's feet is well within acceptable limits. The cross section and robustness
of these feet are minimized to reduce thermal losses:
a more detailed analysis of the structure, leading to possible reinforcement, must
be performed to allow for an even higher inclination range. The operation of the cryostat
with superfluid helium is advantageous: the magnet's end opposite to
the MFB is still immersed in liquid He when in its highest position.

The cryogenic piping and electrical connections to feed the
15 kA needed can easily be made flexible enough to allow a rotation of the girder
about a vertical axis, preferably centered on the MFB, to a range of
$\sim \pm 40^o$. 
\section*{5. Discussion}

The axion-to-photon conversion via the Primakoff effect inside the
LHC model magnets is the established working principle \cite{ps,dml,fh,pvv}
of the proposed axion helioscope.
The same effect allows for axion
production inside celestial objects like our Sun, which 
translates into  exotic 
energy-loss mechanisms that may determine their evolution.
Such an experiment will also be sensitive to the suggested 
monochromatic 14.4 keV nuclear axion emission from the 
Sun's core 
\cite{sm}. 
Besides the
speculated role of relic axions as dark matter constituents and
their potential ability to solve the strong CP problem, the question of their
existence or non-existence becomes of paramount importance with far-reaching
consequences for particle (astro)physics and cosmology.

In conclusion,
compared with the previously achieved experimental results, the proposed CERN experiment has the following
advantages :
~{\bf 1)}~ the active tracking of the Sun is very important because it
increases the alignment period by a factor $\approx$35 (for $\pm1.5^o$ tracking)
 compared with a fixed
magnet geometry (see {\sf Fig. 2a,b}). This period scales  with the
maximum tracking angle along the horizontal and vertical axis.
For comparison, in the suggestive work of
ref. \cite{fh} 
the alignment period is only  $\approx$30 sec per year. ~
~{\bf 2)}~ Solar axion signals
are expected to show up at precisely predicted
time-windows, displaying a semi-diurnal periodicity,  allowing us to exclude systematical
uncertainties and providing a unique signature.
The expected improvement in the experimental coupling constant
$g_{a\gamma \gamma}$ can be as much as a factor 30 to $\sim$ 50 within 1 to 3 years 
for 3$\times$2 months of exposure, depending on the attainable tracking range.
~~{\bf 3)}~Two twin aperture model magnets in series with $\pm~10^o$ vertical tracking
can provide an improvement by a factor as large as $\sim$100.
~~{\bf 4)}~ With the same detection system,
low- as well as high-energy celestial  axions, whatever their origin, can be detected simultaneously due to
the implicit directionality (coming from the Earth's motion),
entering a serendipitous experimental realm, that of {\it axion astronomy}. For this kind of search, no active
tracking is necessary, but is desirable (see {\sf Fig. ~4}).

 Because of the associated  increase in
 the mass scale up to $M\simeq 2.5\cdot 10^{10}$ GeV,
this search is sensitive to potentially new physics
 beyond the Standard Model,
 taking advantage of the LHC-magnet testing program
 and facilities.

\vskip1.5cm

\noindent
{\bf Acknowledgements}

\noindent
We would like to thank Y. Semertzidis (BNL) for many informative discussions.

\newpage

\noindent
{\Large \bf Figure captions}
\vskip1.0cm

\noindent
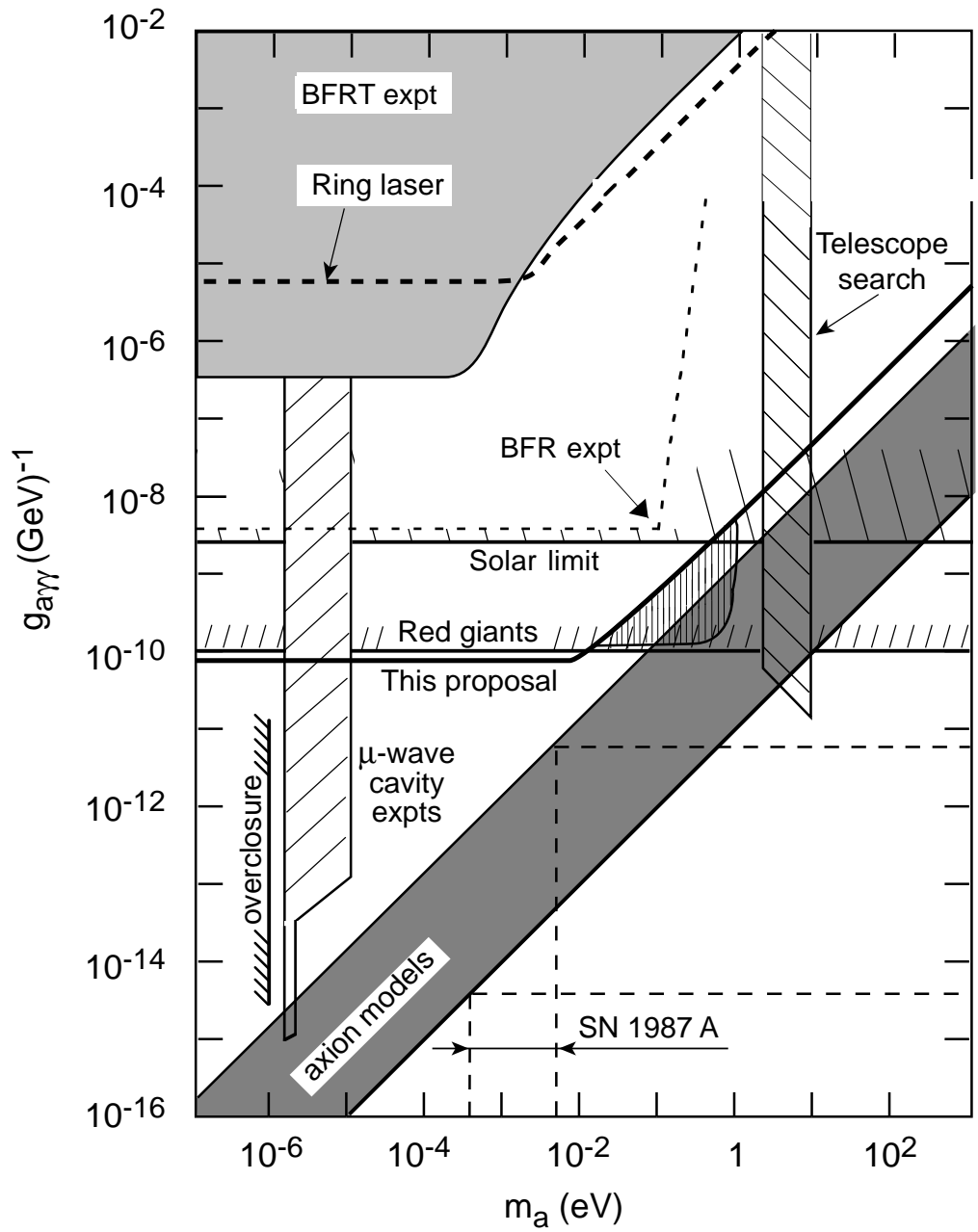
{\bf  Figure~ 1.}

\noindent
Limits on the axion coupling constant to two photons 
($g_{a\gamma\gamma}$) versus its rest mass from laboratory and
astrophysical considerations \cite{yks}. The theoretically-favoured 
relation between the axion-to-photon coupling strength and the 
axion mass is also shown.
 The recent limit from SN1987A is taken from 
ref. \cite{keil}.
The expected limit to be achieved with this proposal 
(L=10 m, $\pm5^o$ vertical tracking, $\pm 40^o$ horizontal) is also shown;
the attached hatched region refers to measurements with He gas in the
magnetic field pipes. The astrophysical limits are largely parameter-dependent
and subject to frequent revisions. References for the BFRT, BFR, telescope, cavity,
solar and red giant limits are \cite{bfrt},\cite{dml},\cite{mtr,kitty},\cite{vanb,ch,io},
\cite{newsol},\cite{redg}, respectively.

\vskip0.5cm

\noindent
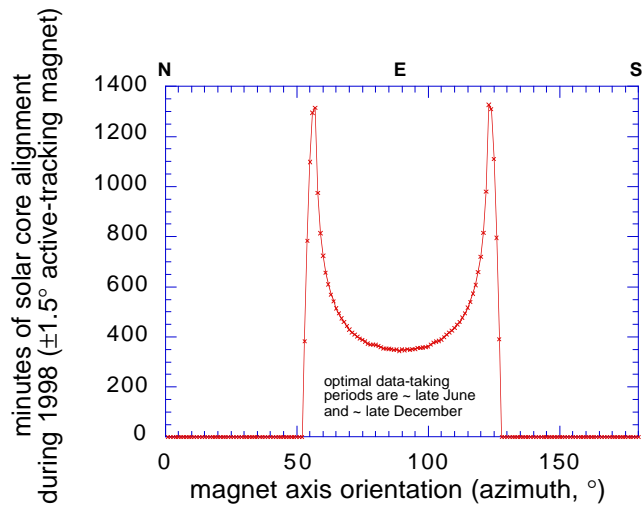
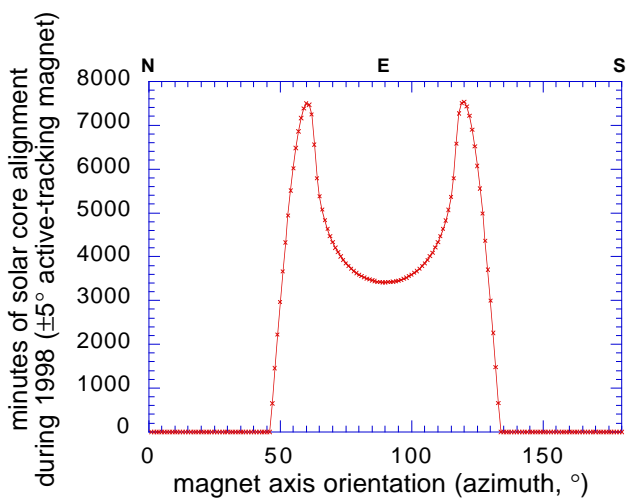
{\bf  Figure~ 2.}

\noindent
 Time of solar alignment during one year for {\sf \it fixed} horizontal magnet orientations  at CERN:
 {\bf (a)}~for a completely stationary magnet,
  {\bf (b)}
  for $\pm~1.5^o$ tracking of the solar core in all directions around the otherwise stationary magnet axis
 orientation, and  {\bf (c)}
 for $\pm 5^o$ tracking in all directions around the otherwise stationary magnet axis
 orientation.

\vskip0.5cm

\noindent
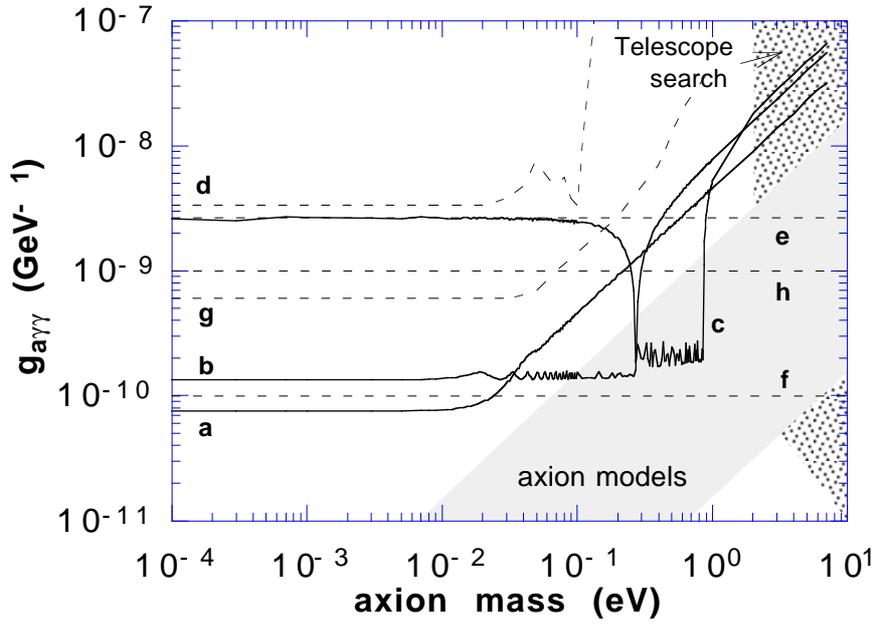
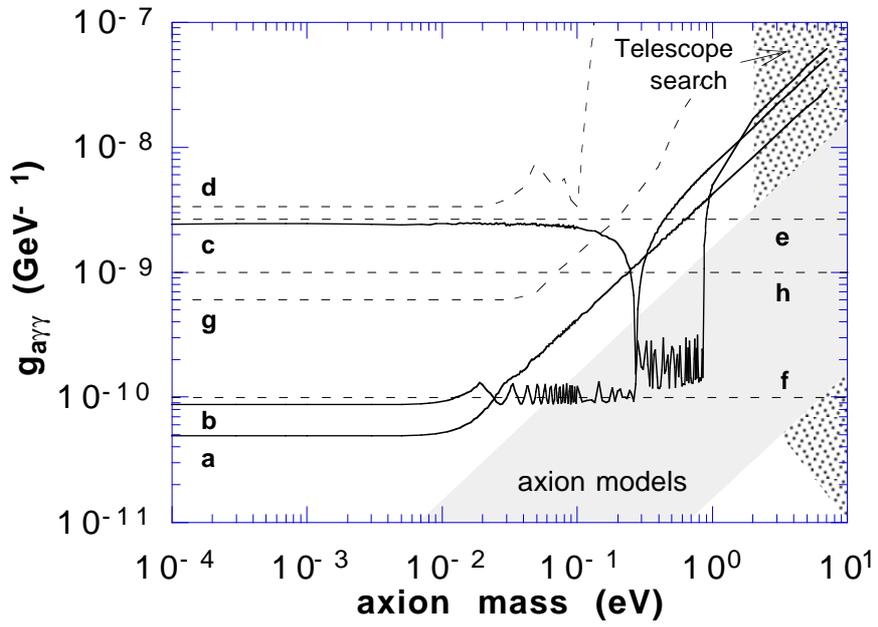
{\bf  Figure~ 3.}

\noindent
 Detailed limits on the coupling strength of axions to two photons
 as a function of the axion rest mass : The top figure corresponds to 
 a 10 m LHC test magnet with $\pm 5^o$ vertical tracking and $\pm 40^o$ horizontal tracking,
 the bottom figure is for two such magnets in series with an increased vertical tracking  
 of $\pm 10^o$. {\bf (a)} After 1 y with vacuum in the pipe line, {\bf (b)} an additional 1 y with He gas
 pressure increased from 0-1 atm in 100 increments, {\bf (c)} an additional 1 y with 1-10 atm in 
 365 increments, {\bf (d)} previous limits from ref. \cite{dml}, {\bf (e)} limits recently imposed
 by members of this collaboration using an underground Ge detector
 \cite{we}, {\bf (f)} the {\sf \it theoretical} red giant limit \cite{redg}, {\bf (g)} 
 recent limits from the Tokio axion helioscope \cite{mino2,mino3} and {\bf (h)}, the new 
 helioseismological constraints on solar axion emission \cite{newsol}.
 The asymptotic behaviour for large axion masses 
 in {\bf (d)} is a conservative approximation \cite{semer}. The "telescope search" limits  
 arise from the absence of an axion-decay quasi-monochromatic photon line from
 galactic clusters \cite{mtr,kitty}.
 
\vskip0.5cm

\noindent
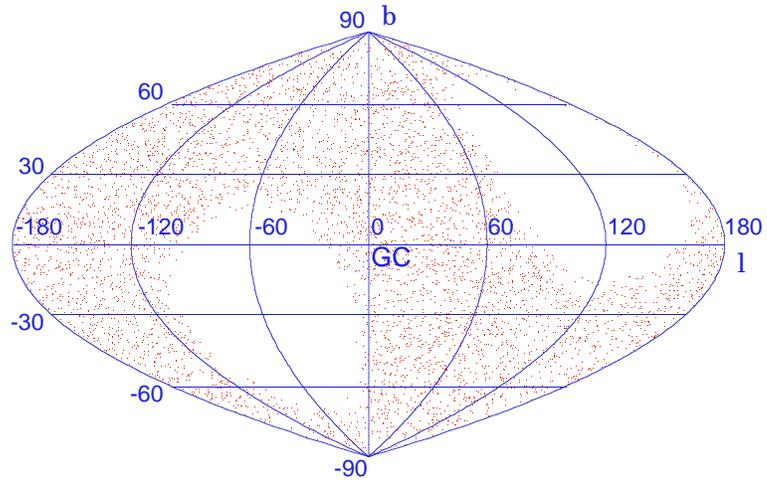
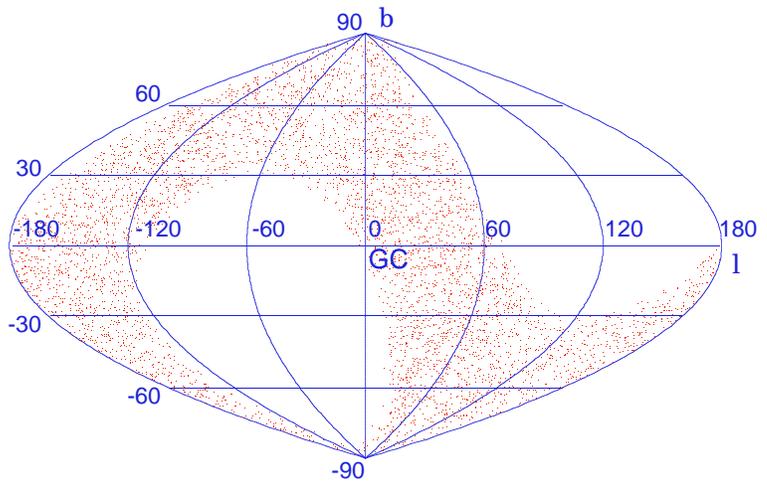
{\bf  Figure~ 4.}

\noindent
 Region of the sky (dotted) in galactic coordinates (the galactic center (GC) is at $b=0^o$, 
 $l=0^o$, the galactic plane at $b=0^o$), scanned by the suggested axion telescope during 
 1998 for {\bf a)} a magnet with $\pm 5^o$ vertical and $180^o$ horizontal tracking 
 and, {\bf b)} the same magnet with a horizontal tracking range limited to NE$-$SE (enough 
 to track sunset and sunrise during the whole year). It is evident that a
  horizontal tracking as large as possible is desirable to explore regions of interest such as the GC. For
   modest vertical tracking, regions of the sky 
   with $\vline \delta \vline > 90^o-\vline \phi \vline$, where $\delta$ is the equatorial 
   declination and $\phi$ is the magnet's geographical latitude ($46.25^o~N$ for CERN), 
   cannot be explored regardless of horizontal tracking ability \cite{fh}. The present epoch
     equatorial coordinates of GC are  $\alpha = 265.5^o$ and $\delta =-28.9^o$,   while 
     the galactic north pole is at $192.2^o$ and $27.4^o$, respectively.

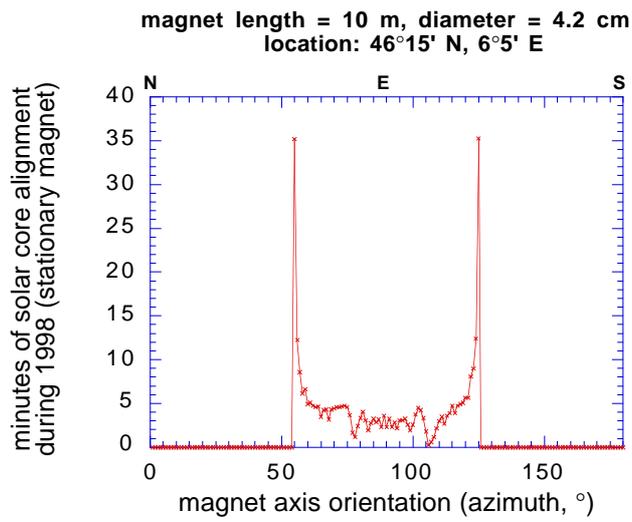
\end{document}